\documentclass[journal]{IEEEtran}
\ifCLASSINFOpdf

\else
  
\fi

\usepackage{cite}
\usepackage{amsfonts}
\usepackage{amsmath}
\usepackage[pdftex]{graphicx}
\usepackage{amsmath,amssymb,amsfonts}
\usepackage{graphicx}
\usepackage{booktabs} %表格线条加粗所需
\usepackage{textcomp}
\usepackage{xcolor}
\usepackage{float}
\usepackage{amsmath, bm}
\usepackage[ruled,linesnumbered,vlined]{algorithm2e}

\usepackage{makecell}

\usepackage{stfloats}

\usepackage{CJK}
\usepackage{indentfirst}
\usepackage{amsmath}
\usepackage{cases}
\usepackage{geometry}

\usepackage{hyperref}
\hypersetup{
            colorlinks=true,
            linkcolor=blue,
            anchorcolor=blue,
            citecolor=blue,
            filecolor=blue,
            urlcolor=blue
            }
\makeatletter
\newcommand{\ccc}[1]{{\color{blue}\cite{#1}}}
\makeatother

\definecolor{b}{rgb}{0,0,0}

\usepackage{amsmath}

\begin{document}

\title{Computation-Efficient Aerial-Marine Integrated Networks for Search and Rescue via Cooperative HAPS, UAVs and MASSs}

 \author{Zhen~Wang,~\IEEEmembership{Student Member,~IEEE,}
         Bin~Lin,~\IEEEmembership{Senior Member,~IEEE,}
        and~Qiang (John) Ye,~\IEEEmembership{Senior Member,~IEEE}
%         Yuguang~Fang,~\IEEEmembership{Fellow,~IEEE}
%         and~Xiaoling~Han,~\IEEEmembership{Student Member,~IEEE}% <-this % stops a space
% \thanks{Copyright (c) 20xx IEEE. Personal use of this material is permitted. However, permission to use this material for any other purposes must be obtained from the IEEE by sending a request to pubs-permissions@ieee.org.}
\thanks{\emph{(Corresponding author: Bin Lin.) }}
\thanks{Zhen Wang is with the Information Science and Technology College, Dalian Maritime University, Dalian, 116026, China, and Communication Engineering of Dalian Neusoft University of Information, Dalian, 116023, China. E-mail: wangzhen\underline{~}jsj@neusoft.edu.cn.}% <-this % stops a space
\thanks{Bin Lin is with the Information Science and Technology College, Dalian Maritime University, Dalian, 116026, China. E-mail: binlin@dlmu.edu.cn.}% <-this % stops a space
\thanks{Qiang (John) Ye is with the Department of Electrical and Software Engineering, Schulich School of Engineering, University of Calgary, 2500 University Drive NW, Calgary, AB T2N 1N4. Email: qiang.ye@ucalgary.ca.}% <-this % stops a space
\thanks{This manuscript has been accepted by IEEE Transactions on Cognitive Communications and Networking, DOI:10.1109/TCCN.2025.3642113.}

% <-this % stops a space

 }

% make the title area
\maketitle

% As a general rule, do not put math, special symbols or citations
% in the abstract or keywords.
\begin{abstract}
In this paper, we propose an aerial-marine integrated network architecture tailored for maritime search and rescue (MSAR) operations, where a high altitude platform station (HAPS) and maritime autonomous surface ships (MASSs) are equipped with edge servers, providing computing services for surveillance unmanned aerial vehicles (UAVs). 
These UAVs generate substantial volumes of computation-intensive and delay sensitive tasks, some of which can be strategically offloaded to both MASSs and HAPSs for distributed processing.
We formulate the joint task offloading and resource allocation as a mixed-integer nonlinear program (MINLP) to minimize the system computation overhead (CO), quantified by a weighted sum of task completion time and energy consumption. Given the stochastic nature of task arrivals from UAVs and the dynamic characteristics of marine channel conditions, we employ a two-stage optimization to decompose the problem into four tractable subproblems, i.e., (1) edge server selection, (2) transmission power allocation, (3) edge computing resource allocation, and (4) local computation resource allocation. 
In stage I, we optimize the edge server selection via many-to-one matching. In stage II, we determine the transmission power of each UAV with quasi-convex optimization and solve the edge computing and local computing resource allocation subproblems via a projected gradient descent method and convex optimization, respectively. 
Simulation results validate that the proposed scheme can achieve superior system performance over benchmark schemes.

\end{abstract}

% Note that keywords are not normally used for peerreview papers.

\IEEEpeerreviewmaketitle

\section{Introduction}

\IEEEPARstart{M}{aritime} search and rescue (MSAR) operations are critical for ensuring safety and minimizing casualties in emergency situations at sea\ccc{10499870},\ccc{8916645}. It is paramount to perform timely search-and-rescue missions within the golden window of post-disaster rescue. In recent years, MSAR operations have faced increasing challenges due to the expanding scope of maritime activities, the complexity of rescue missions, and the vast and unpredictable nature of maritime environments\ccc{9014520},\ccc{9800919}. Traditional MSAR methods, which rely heavily on human resources and terrestrial networks, are often limited by geographical constraints, weather conditions, and availability of rescue assets\ccc{10490255}. In general, an effective MSAR operation necessitates an immediate response, including swift network deployment, real-time information collection, and low-latency data processing\ccc{10757774}. Nevertheless, the scarcity of conventional maritime communication and computing resources has posed potential obstacles to rescue operations\ccc{10817534},\ccc{ZHANG2025100275},\ccc{Zhang002025}. To address these challenges, there is a growing interest in leveraging advanced technologies to enhance the efficiency and effectiveness of MSAR operations\ccc{9079822}. 

\textcolor{b}{Multi-access edge computing (MEC) has emerged as a transformative technology that brings computational resources closer to the network edge, enabling low-latency and high-efficiency task processing\ccc{8419289},\ccc{10841367},\ccc{10018447},\ccc{8344453},\ccc{9454395},\ccc{10771784},\ccc{10772404},\ccc{KHAN2025100317},\ccc{Wang2025},\ccc{Zhang012025}. The integration of MEC into maritime networks can significantly enhance the network capability to handle computation-intensive and delay-sensitive tasks. Given the inherently limited maritime resources, most existing maritime MEC studies have established network models supported by shore-based, air-based, or space-based infrastructures as the communication foundation for optimizing maritime MEC network performance\ccc{10490255},\ccc{10817645},\ccc{10829786}. 
Qi \emph{et al.} proposed an aerial-assisted maritime rescue system employing multiple surveillance UAVs and a relay UAV for disaster data acquisition in\ccc{10490255}. The study jointly optimizes task offloading decisions, relay UAV deployment, and surveillance UAV-rescue target associations to minimize total system latency.
In\ccc{10817645}, Zhang \emph{et al.} proposed a shore-based edge computing network architecture, where an unmanned surface vessel (USV) equipped with edge servers provides computing services for offshore users and can offload partial tasks to shore-based base stations (BSs) for processing.}
A three-tier edge computing network architecture for maritime devices is proposed in\ccc{10829786}, where computational tasks generated by marine devices can be either processed locally, offloaded to satellites, or further relayed to shore-based BSs for further execution.

In the context of MSAR, the dynamic and resource-constrained nature of maritime environments necessitates efficient task offloading and resource allocation strategies to optimize the performance of maritime networks for MSAR\ccc{10040976}. The integration of aerial computing platforms, i.e., unmanned aerial vehicles (UAVs) and high altitude platform station (HAPS) systems into MSAR operations has emerged as a promising solution, offering enhanced mobility, flexibility, and coverage\ccc{10726729},\ccc{9772280},\ccc{10437686}. 
UAVs are highly practical for deployment in disaster-affected areas due to their exceptional maneuverability, flexible deployment capabilities, reliable line-of-sight (LoS) communication, and cost efficiency\ccc{10234672},\ccc{9628162}. They play a critical role in supporting rescue missions, including disaster area monitoring, real-time data collection, and aerial search and rescue operations\ccc{10365400}.  However, the limited onboard battery capacity and computing capabilities of UAVs can significantly restrict their endurance and efficiency when performing computation-intensive tasks required for rescue missions. 

Recently, HAPS systems have emerged as key enablers for six generation (6G) networks\ccc{10529611},\ccc{9651548},\ccc{9749222}, offering unique capabilities as stratospheric wireless communication stations operating at altitudes of around 20 km. These aerial platforms, such as airships or balloons, can provide a bird's-eye line-of-sight (LoS) view over extensive ground areas, enabling wide coverage with a radius of 50-500 km\ccc{10103832}. With their large payload capacity (typically $\geq$ 100 kg), HAPS can be equipped with advanced resources, i.e., large batteries, antennas, and computation and storage resources, making them powerful supplements for maritime MEC\ccc{9815183}. Considering the limitations of UAVs in terms of energy and computational prowess, a practical approach to enhance MEC efficiency entails allocating UAVs' tasks to both HAPS and sea surface stations for subsequent processing. 

However, establishing UAV-HAPS-assisted maritime MEC for MSAR faces technical challenges. First, the dynamic nature of task generation from UAVs, coupled with the ever-fluctuating communication quality of maritime wireless channels, are susceptible to various marine environmental influences. Second, the aerial-marine integrated network comprises a multitude of computation options, making it particularly challenging to select suitable computation equipment and optimally utilize the available resources\ccc{Huang2025}. Third, there are multidimensional resources in aerial-marine integrated network, such as the computation resources of HAPS and sea surface stations, and the transmission power of UAVs. Efficiently allocating these resources and maximizing their utilization to enhance network performance poses a formidable challenge. 

To overcome the aforementioned challenges, we propose a novel aerial-marine integrated network architecture for MSAR and introduce an approach for joint optimization of task offloading and resource allocation to minimize the system computation overhead (\textcolor{blue}{CO}). 
The main contributions are summarized as follows:

\begin{itemize}
\item[$\bullet$] 
\emph{System Architecture}: We delve into the exploration of an aerial-marine integrated network architecture for MSAR, wherein multiple surveillance UAVs installed with lightweight computing resources are dispatched to a disaster area for an MSAR mission. A HAPS equipped with extensive computing resources is deployed to provide aerial edge capability to the UAVs. Concurrently, numerous maritime autonomous surface ships (MASSs) equipped with edge servers are deployed to the disaster areas to assist UAVs in processing tasks at closer distances. The computation-intensive tasks of UAVs can be offloaded to the HAPS or an MASS via multi-access approach. 
\item[$\bullet$] 
\emph{Problem Formulation}: We propose a novel joint optimization framework for task offloading and resource allocation, formulated as a mixed-integer nonlinear program (MINLP). The objective is to minimize the system \textcolor{blue}{CO}, defined theoretically as a weighted-sum of task completion time and energy consumption. 
\item[$\bullet$] 
\emph{Algorithm Design}: Given the complexity of directly solving the formulated problem, we employ a two-stage optimization and decompose the primal problem into four subproblems: 1) sever selection; 2) transmission power allocation; 3) edge computing resource allocation; 4) local computing resource allocation. The first subproblem is addressed using a many-to-one matching approach, while the second subproblem is solved through quasi-convex optimization techniques. For the third subproblem, we apply a projected gradient descent method, and the fourth subproblem is resolved using convex optimization. On this basis, we introduce a joint computation offloading and resource allocation (JCORA) algorithm to optimize the total \textcolor{blue}{CO} of the system. 
\item[$\bullet$] 
\emph{Performance Evaluation}: We conduct extensive simulations to validate the effectiveness of the proposed solution. The results demonstrate that our proposed algorithms can minimize the \textcolor{blue}{CO} of the system and improve the efficiency in comparison with the benchmark schemes.
\end{itemize}

The remainder of this paper is organized as follows. Section II summarizes the related work. The system model is presented in Section III. The problem formulation is given in Section IV. Section V presents the problem transformation and solution design. Section VI provides the performance evaluation results, and Section VII concludes the paper.
% Table I lists the key notations according to the order in which they first appear in the paper.

% \begin{table}[htbp]
% \centering
% \caption{List of Key Notations}
% \begin{tabular}{c|c}%l=left, r=right,c=center分别代表左对齐，右对齐和居中，字母的个数代表列数
% \hline
% \hline
% $\mathbf{Symbol}$ & $\mathbf{Meaning}$  \\ 
% \hline
% $M_k$  &The numbedr of TUs under MIS $k$ \\
% \hline
% $p_{k,i}$  &Processing power of MIS $k$ for TU $i$\\
% \hline
% \hline
% \end{tabular}
% \end{table}

\section{Related Work}
In this section, we review existing research works related to aerial MEC and marine MEC approaches, respectively. The differences between the related works and our work is summarized in Table I.

\begin{table*}[h!]\color{b}
	
	\begin{center}
		\caption{Comparison Between Related Work and our work.}
		\resizebox{\textwidth}{!}{
			\begin{tabular}{|c|c|c|c|c|c|c|c|} % <-- Alignments: 1st column left, 2nd middle and 3rd right, with vertical lines in between
				\hline
				{References} & {\makecell[c]{Search and \\ rescue scenario}} & {\makecell[c]{Offloading route\\ (UAV$\to$ HAPS)}} &  {\makecell[c]{Offloading route\\ (UAV$\to$ MASS)}}
		    	& {\makecell[c]{Optimizing\\ latency}}
		    	& {\makecell[c]{Optimizing energy\\ consumption}}
		    	& {\makecell[c]{Task\\ offloading}}
		    	& {\makecell[c]{Resource\\ allocation}}		
		    
		    	\\		
				\hline
				\ccc{10499870}&\checkmark  &$\times$  &$\times$  &\checkmark  &\checkmark  &\checkmark  &\checkmark \\
				\hline
				\ccc{10490255}&\checkmark  &$\times$  &$\times$  &\checkmark  &$\times$  &\checkmark  &$\times$ \\
				\hline
				\ccc{10726729} &\checkmark  &$\times$  &$\times$  & \checkmark &$\times$  &\checkmark  &$\times$ \\	
				\hline
				\ccc{9780606} &$\times$  &$\times$  &$\times$  &$\times$  &$\times$  &\checkmark  &\checkmark \\
				\hline
				\ccc{10546306} &$\times$  &$\times$  &$\times$  &\checkmark  &$\times$  &\checkmark  &$\times$ \\
				\hline
				\ccc{10382630} &\checkmark &$\times$ &$\times$ &\checkmark &\checkmark &\checkmark &\checkmark \\
				\hline
				\ccc{9672750} &$\times$  &$\times$  &$\times$  &$\times$  &\checkmark  &\checkmark  &\checkmark \\
				\hline
				\ccc{10736665} &$\times$  &$\times$  &$\times$  &$\times$  &$\times$  &\checkmark  &\checkmark \\
				\hline
				\ccc{9954169} &$\times$  &$\times$  &$\times$  &$\times$  &$\times$  &\checkmark  &\checkmark \\
				\hline
				\ccc{9515574} &$\times$  &$\times$  &$\times$  &$\times$  &\checkmark  &\checkmark  &\checkmark \\
				\hline
				\ccc{9743399} &$\times$  &$\times$  &$\times$  &$\times$  &\checkmark  &\checkmark  &\checkmark  \\
				\hline
				\ccc{9996964} &$\times$  &$\times$  &$\times$  &$\times$  &\checkmark  &\checkmark  &\checkmark \\ 
				\hline
				\ccc{10679218} &$\times$  &$\times$  &$\times$  &$\times$  &\checkmark  &\checkmark  &\checkmark \\
				\hline
				\ccc{10233325} &$\times$  &$\times$  &$\times$  &$\times$  &$\times$  &\checkmark  &\checkmark \\
				\hline
				\ccc{9714482} &$\times$  &\checkmark  &$\times$  &$\times$  &$\times$  &\checkmark  &$\times$ \\
				\hline
				\ccc{10449425} &$\times$  &$\times$  &\checkmark  &$\times$  &\checkmark  &\checkmark  &\checkmark \\
				\hline
				\ccc{10296008}&$\times$  &$\times$  &$\times$  &$\times$  &\checkmark  &\checkmark  &$\times$ \\
				\hline
				\ccc{2024AUV} &$\times$  &$\times$  &$\times$  &$\times$  &\checkmark  &\checkmark  &$\times$ \\
				\hline
				\ccc{10518032} &$\times$ &$\times$  &$\times$  &$\times$  &\checkmark  &\checkmark  &\checkmark   \\
				\hline
				\ccc{10376220} &\checkmark &$\times$  &$\times$  &\checkmark  &\checkmark  &\checkmark  &\checkmark   \\
				\hline
				\ccc{10224351} &$\times$  &$\times$  &$\times$  &\checkmark  &$\times$  &\checkmark  &\checkmark \\
				\hline
				\ccc{11006480} &$\times$  &\checkmark  &$\times$  &$\times$  &\checkmark  &\checkmark  &$\times$ \\
				\hline
				\ccc{11045994} &\checkmark  &$\times$  &\checkmark  &\checkmark  &$\times$  &\checkmark  &\checkmark \\
				\hline
				Our work &\checkmark  & \checkmark &\checkmark  &\checkmark  & \checkmark &\checkmark  &\checkmark \\
				\hline
				
		\end{tabular}	}
	\end{center}
	
\end{table*}

\subsection{Aerial MEC}
Aerial MEC mainly explores the integration of UAVs, HAPS, and other aerial platforms to enhance edge computing capabilities in challenging environments\ccc{10499870},\ccc{10726729},\ccc{9780606},\ccc{10546306}. Researchers have focused on leveraging the mobility, flexibility, and LoS advantages of aerial platforms to provide computational and communication support in areas with limited terrestrial infrastructure, such as remote regions, disaster zones, and maritime environments. Some studies have demonstrated the potential of UAVs as flying edge servers, offering low-latency processing and data offloading for terminal devices\ccc{10382630},\ccc{9672750},\ccc{10736665},\ccc{9954169}. 
For instance, Liu \emph{et al.} proposed a multiple input single output (MISO) UAV-assisted MEC network framework to tackle the challenge of degraded channel quality in traditional MEC networks caused by multipath propagation and blockages\ccc{9672750}. 
Zhao \emph{et al.}  investigated the deployment of UAVs to collaboratively provide edge computing services, jointly optimizing UAV trajectories and task management strategies to maximize the system throughput\ccc{10736665}.
The HAPSs have also been explored as stable, long-endurance platforms for aerial MEC, capable of covering large areas and supporting heavy computational workloads\ccc{9515574},\ccc{9743399},\ccc{9996964},\ccc{10679218},\ccc{10233325},\ccc{9714482}. 
For instance, Ding \emph{et al.} investigated a satellite-aerial integrated edge computing network to combine a low-earth-orbit(LEO) satellite and HAPSs to provide edge computing services for ground user equipment\ccc{9515574}.
Lakew \emph{et al.} proposed a heterogeneous aerial access Internet of Things (IoT) network comprising a HAPS and multiple UAVs with computation and communication capabilities to provide task offloading services for the resource-limited IoT Devices\ccc{9743399}. \textcolor{b}{Jia \emph{et al.} proposed a hierarchical model composed of an HAP and multi-UAVs in\ccc{11006480} to minimize the total energy cost with the chance constraint for remote  ground users.}

\subsection{Marine MEC}
Marine MEC mainly focuses on extending edge computing capabilities to maritime environments, where traditional infrastructure is often unavailable or impractical.
Researchers have explored the integration of unmanned surface vehicles (USVs)\ccc{10449425}, UAVs\ccc{10296008}, HAPSs\ccc{9996964}, autonomous underwater vehicles (AUVs)\ccc{2024AUV}, LEO satellites\ccc{10518032} and other edge platforms to address the unique challenges of maritime operations, such as search and rescue, environmental monitoring, and vessel traffic management. 
In\ccc{10449425}, Zeng \emph{et al.} proposed an energy-efficient USV fleets-assisted collaborative computation offloading scheme for smart maritime services, where the USVs were equipped with computing resources to provide edge services for the UAVs.
In\ccc{10376220}, Lyu \emph{et al.} formulated a joint task offloading and resource allocation problem for a UAV emergency communication scenario, aiming to minimize the \textcolor{blue}{CO} of the terminal device.
In\ccc{9996964}, Li \emph{et al.} exploited the secure computation
offloading with the assistance of a HAPS to perform computation offloading and provide cooperative jamming for USVs.
Dai \emph{et al.} proposed a hybrid offshore and aerial-based MEC  network to enhance the offloading efficiency and minimize the maximum workloads latency of marine devices\ccc{10224351}. 
\textcolor{b}{You \emph{et al.} proposed a cooperative MEC framework composed of UAVs and vessels in\ccc{11045994} to minimize the total execution time, with consideration of uncertain MIoT tasks.}

Although many existing studies focus on aerial MEC and marine MEC, the utilization of both HAPS and MASSs to create multiple layers of edge computing services for UAVs, especially for MSAR operations, need further investigation. In this paper, leveraging the high maneuverability and flexible deployment of UAVs, we utilize them as surveillance devices for MSAR and propose an innovative aerial-marine integrated network architecture, which enables the UAVs to offload computation-intensive tasks to either HAPS or MASSs for further processing.

\section{System Model}
This section first introduces an aerial-marine integrated network, where the UAVs can offload their tasks to an MASS or a HAPS edge for further processing. Then, the communication and computing models are presented.

\subsection{Network Model}
We consider an aerial-marine integrated network with MEC functionality, consisting of one HAP and multiple MASSs, cooperatively supporting task offloading
for connected UAVs, as shown in Fig. 1.
In the considered scenario, a group of UAVs equipped with cameras are distributed to collect data from the disaster areas according to the pre-set trajectories. Meanwhile, an HAP and a certain number of MASSs equipped with edge servers are dispatched to receive the critical data collected by UAVs. 
Moreover, each MASS is assigned a dedicated service area that maintains non-overlapping boundaries with adjacent MASSs. Additionally, all UAVs maintains a constant altitude of $H_0$ to minimize unnecessary energy expenditure associated with frequent altitude adjustments in response to potential obstacles\ccc{10382630}.
We posit a one-to-one correspondence between HAPS and MASSs with edge servers.  
The set of UAVs and edge servers are denoted as $\mathcal{I}=\left\{1, 2,..., i,..., I\right\}$ and $\mathcal{M}=\left\{0,1, 2,..., m,..., M\right\}$, respectively. Therefore, there are $M + 1$ edge servers available to serve the UAVs. For convenience, we denote the $m$-th edge server as server $m$, where $m = 0$ indicates the edge server connected with the HAP, and others mean edge servers connected with the MASSs. Each edge server is designed to provide communication and computing services for UAVs under its coverage, the number of which is denoted by  $\Lambda_m$ ($0\leq\Lambda_m\leq\Lambda_m^{\max}$).

\begin{figure}[htbp]
	\centerline{\includegraphics[width=0.45\textwidth]{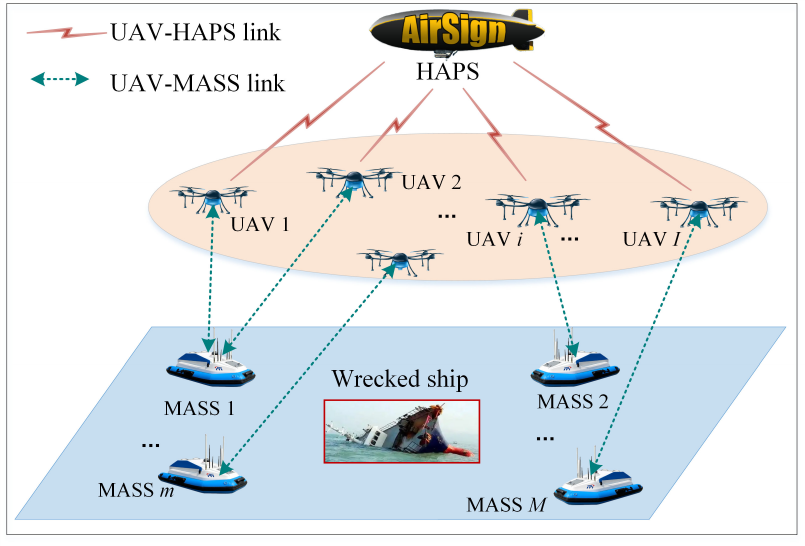}}
	\caption{Network model.}
	\label{fig}
\end{figure}

Typically, UAVs generate a significant volume of computation-intensive tasks at the onset of each time slot, while being constrained by limited onboard computational resources and battery capacity. To address these limitations, UAVs can leverage the computational capabilities of either HAPS or MASS platforms through task offloading mechanisms. 
The server selection decision for UAV $i$ is defined as $s_i^m$, where $s_i^m=1$ indicates that UAV $i$ selects server $m$ to handle tasks in the time slot $t$, or else $s_i^m=0$. The set $\bm{s}=\left\lbrace s_i^m|i\in\mathcal{I}\right\rbrace $ indicates the server selection decisions of all UAVs. In addition, the transmission power of UAV $i$ during communication with an edge server is represented as $p_i$, and $0\leq p_i\leq p_i^{\max}$. The set $\bm{p}=\left\lbrace p_i|i\in\mathcal{I}\right\rbrace $ indicates the transmission power of all UAVs.

Let $a_i$ denote the number of task arrivals at UAV $i$, which is randomly drawn from $\left[0,a^{\max}\right] $ to capture the temporal variations in task arrivals. $a^{\max}$ is the upper bound of the number of arriving tasks. The UAVs may request for different types of tasks which vary in input data size and required CPU cycles. To simplify the system model, we assume that each task has the same size of $D$ (in bits) and the expected number of CPU cycles required by one bit is $\alpha$\ccc{10453323}. Additionally, all the tasks generated by UAV $i$ must be completed within its latency requirement $T_i^{th}$. 

The three-dimensional Cartesian coordinate system is employed to represent the position coordinates of the HAPS, UAVs and MASSs. Generally, HAPSs can endure at a fixed position around the altitude of $20$ km for several months, which can serve as stable BSs in the air due to large coverage and powerful payloads\ccc{10373079}. The position of an HAPS is expressed as $\textbf{q}_{HAP}=\left(x_{H},y_{H},z_{H}\right)$.
The position of UAV $i$ is denoted as $\textbf{q}_i=\left(x_i,y_i,H_0\right)$.  
Considering the safety distance between the UAVs, the position of each UAV should satisfy $\vert\vert\textbf{q}_i-\textbf{q}_j\vert\vert\geq d^{\min}.$
Similarly, the position of MASS $m$ is denoted as $\textbf{q}_m=\left(x_m,y_m,z_m\right)$.

\subsection{Communication Model}
The communication model consists of UAV-to-MASS communication and UAV-to-HAPS communication. The orthogonal frequency-division multiple access (OFDMA) protocol is employed for each MASS or the HAP
channel access.

\emph{(1) UAV-to-MASS Communication Model}

Given the relatively open offshore environment, characterized by stronger direct signals, and considering that the primary factors influencing overseas radio channels are the multipath effect due to sea volatility and the impact of extreme weather, we regard the communication between UAVs and MASSs as air-to-sea channels, which exhibit Rician fading\ccc{8960465}. This fading can be viewed as a composite of large-scale and small-scale fading\ccc{8770592}. 

The large-scale path loss model is expressed as\ccc{8528349}
\begin{equation}
	L_i^M \mid_{dB}=L_0+10\lambda\log_{10}\left(\frac{d_i^m}{d_0} \right)+X_{\sigma}+\varrho F ,
\end{equation}
where $d_i^m=\sqrt{(x_m-x_i)^2+(y_m-y_i)^2+H_0^2}$ denotes the distance between UAV $i$ and server $m$ ($m=1,2,...,M$), $L_0$ is the path loss at the reference distance $d_0$, $\lambda$ denotes the path-loss exponent which could be reduced to less than 2 due to the ducting effect over the sea surface\ccc{8528349}, $X_{\sigma}$ describes the shadow fading caused by, e.g., sea waves under high sea state conditions, $F$ is an adjustment parameter to  incorporate the fast moving of UAVs, and the value of $\varrho$ is set to $1$ or $-1$ depending on the moving direction of the UAVs (towards or away from the ground site)\ccc{10966051}.
 
The small-scale Rician fading $\tilde{u}_i^m$ is represented as
\begin{equation}
	\tilde{u}_i^m=\sqrt{\frac{K_R}{1+K_R}}+\sqrt{\frac{1}{1+K_R}}g_i^m,
\end{equation}
where $g_i^m \in \mathcal{CN}(0,1)$ and $K_R$ is the Rician factor.
On the ocean, MASSs typically navigate along established shipping routes, from which historical or pre-measured data can be obtained. Leveraging this derived data, we establish a correlation between specific locations and channel state information (CSI). With this correlation, we can determine the corresponding CSI for each location. Hence, we postulate that information on path loss and the Rician factor is accessible\ccc{8770592}.

Then, the channel coefficient is expressed as
\begin{equation}
	h_i^m=\left(L_i^M\right)^{-1/2} \tilde{u}_i^m,
\end{equation}

The channel gain between UAV $i$ and server $m$ ($m=1,2,...,M$) is expressed as
\begin{equation}
H_i^m=G^UG^M\mid h_i^m\mid^2 , m=1,2,..,M,
\end{equation}
where $G^U$ and $G^M$ are the antenna gain of UAVs and MASSs, respectively.

\emph{(2) UAV-to-HAPS Communication Model}

In the UAV-HAPS communication model, each UAV communicates with one HAPS, which hovers at a fixed height in the stratosphere. According to\ccc{9351533}, the path loss when UAV $i$ communicates with the HAPS is calculated as
\begin{align}
L_i^H&=20\log_{10}\left(\frac{4\pi f_id_i^{H}}{c}\right)+\gamma_i^{LoS}\delta_i^{LoS}\nonumber\\
&+\left(1-\gamma_i^{LoS}\right)\delta_i^{NLoS}, 
\end{align}
where $f_i$ denotes the carrier frequency, $c$ is the speed of light,  $d_i^H=\sqrt{(x_H-x_i)^2+(y_H-y_i)^2+(z_H-H_0)^2}$ means the distance between UAV $i$ and the HAP,  $\delta_i^{LoS}$ and $\delta_i^{NLoS}$ denote LoS and non-LoS (NLoS) link loss when UAV $i$ communicates with the HAPS, respectively. $\gamma_i^{LoS}$ denotes the LoS communication probability, calculated as
\begin{equation}
 \gamma_i^{LoS}=\frac{1}{1+\chi_1\exp\left[-\chi_2\left(\varphi_i-\chi_1\right) \right] }
\end{equation}
where $\varphi_i=\tan^{-1}\left[ \left(z_H-H_0\right)/\sqrt{(x_H-x_i)^2+(y_H-y_i)^2}\right] $ means the elevation angle of UAV $i$ to the HAPS, and the values of $\chi_1$ and $\chi_2$ are determined by the state of the environment.
Without loss of generality, the channel gain when UAV $i$ communicates with the HAPS is expressed as\ccc{10373079}
\begin{equation}
	H_i^m=10^{-L_i^H/10}, m=0.
\end{equation}
In summary, according to Shannon theorem, the transmission rate (link capacity) between UAV $i$ and edge server $m$ is calculated by
\begin{equation}
R_i^m=s_i^mW_i^m\log_2\left(1+\frac{p_iH_i^m}{\sigma^2}\right), m=0,1,...,M
\end{equation}
where $W_i^m$ denotes the communication bandwidth allocated to UAV $i$.

\subsection{Computing Model}

\emph{(1) Local Computing Model}

When the server selection decision $s_i^m=0$, UAV $i$ decides to  process $a_i$ tasks locally. We define $\rho_i$ as the computation capacity of UAV $i$, which satisfies $0\leq \rho_i \leq \rho_i^{\max}$. Thus, the computation completion time of $a_i$ tasks is expressed as
\begin{equation}
	t_i^{loc}=\frac{a_iD\alpha}{\rho_i}.
\end{equation}
The energy consumed by local computation hinges on the integrated chip architecture of the UAVs\ccc{10179258}. Then, the energy consumed by local computation of UAV $i$ is calculated as
\begin{equation}\color{b}
	E_i^{loc}=\varepsilon_i\left(\rho_i\right) ^2a_iD\alpha,
\end{equation}
where $\varepsilon_i$ means the effective switched capacitance\ccc{10179258}.

\emph{(2) Edge Computing Model}

When the server selection decision $s_i^m=1$, UAV $i$ decides to offload $a_i$ tasks to an edge server for further processing.
The process of task offloading typically comprises three distinct phases: 1) UAV $i$ conveys the tasks to an edge server via a OFDMA-based uplink; 2) the edge server acknowledges receipt of the computational tasks and allocates the necessary computational resources for task execution; 3) once the tasks have been successfully computed, the edge server transmits the result back to UAV $i$. Since the output of a computational task is generally significantly smaller in size compared to its input\ccc{10376220},\ccc{10453323}, we omit the third phase.

The edge computing completion time $t_i^{edge}$ of UAV $i$ consists of two components, expressed as
\begin{equation}
	t_i^{edge}=t_i^{off}+t_i^{com},
\end{equation}
where $t_i^{off}$ and $t_i^{com}$ denote the task offloading time and the task processing time, respectively.
$t_i^{off}$ mainly depends on the size of tasks and the transmission rate, calculated as
\begin{equation}
	t_i^{off}=\frac{a_iD}{R_i^m}.
\end{equation}
The corresponding energy dissipation required for offloading the computation tasks of UAV $i$ is calculated as
\begin{equation}
	E_i^{off}=p_it_i^{off}=\frac{p_ia_iD}{R_i^m}.
\end{equation}
$t_i^{com}$ mainly depends on the the computation resource allocated to UAV $i$. We denote the computation capacity of edge server $m$ (in the unit of CPU cycles per second) as $F_m$, and the ratio of computing resources allocated to UAV $i$ as $\beta_i^m \in \left[0,1\right]$, satisfying $\sum_{i\in \mathcal{I}}\beta_i^m\leq 1$.  
When $a_i$ tasks of UAV $i$ are offloaded to edge server $m$, the task execution time is calculated as
\begin{equation}
	t_i^{com}=\frac{a_iD\alpha}{\beta_i^mF_m}.
\end{equation}
The corresponding energy dissipation required for processing the computation offloading of UAV $i$ at edge server $m$ is given by
\begin{equation}
	E_i^{com}=\varepsilon_m\left(\beta_i^mF_m\right) ^2a_iD\alpha,
\end{equation}
where $\varepsilon_m$ is the effective switched capacitance of edge server $m$.
Then, Therefore, the total energy consumption of UAV $i$ for edge computing is expressed as
\begin{equation}
	E_i^{edge}=E_i^{off}+E_i^{com}=\frac{p_ia_iD}{R_i^m}+\varepsilon_m\left(\beta_i^mF_m\right) ^2a_iD\alpha.
\end{equation}

\section{Problem formulation}
In an aerial-marine integrated MEC scenario, the \textcolor{blue}{CO} of each UAV is mainly characterized by the task completion time $T_i$ and energy consumption $E_i$, obtained as
\begin{equation}
	T_i=s_i^mt_i^{edge}+\left(1-s_i^m\right)t_i^{loc}
\end{equation} 
and
\begin{equation}
	E_i=s_i^mE_i^{edge}+\left(1-s_i^m\right)E_i^{loc},
\end{equation}
respectively. Our goal is to minimize the \textcolor{blue}{CO} of all the UAVs. Considering the preferences of different UAVs, we define the \textcolor{blue}{CO} function of UAV $i$ as 
\begin{equation}
	\mathcal{G}_i=\mu_i^tT_i+\mu_i^eE_i,
\end{equation}
where $\mu_i^t,\mu_i^e\in \left[0,1\right]$, with $\mu_i^t+\mu_i^e=1, \forall i\in \mathcal{I}$, indicate UAV $i$’s preference on task completion time and energy consumption, respectively, which are determined by the residual battery energy of the UAV itself and the requirement for task completion time. 
For instance, an UAV experiencing short battery life can opt to increase $\mu_i^e$ while decreasing $\mu_i^t$, thereby conserving more energy at the cost of extending the task completion time. 

Based on the system model described in Section III, we aim to minimize the \textcolor{blue}{CO} of all the UAVs (denoted as $\mathcal{G}=\sum_{i\in\mathcal{I}}\mathcal{G}_i$) by jointly optimizing the sever selection decision matrix of UAVs $\bm{s}=\left\{s_i^m\right\}_{i\in\mathcal{I},m\in\mathcal{M}}$, the power allocation matrix of UAVs $\bm{p}=\left\{p_i\right\}_{i\in\mathcal{I}}$, the computing resource allocation matrix of edge servers $\boldsymbol{\beta}=\left\{\beta_i^m\right\}_{i\in\mathcal{I},m\in\mathcal{M}}$, and the local computing resource allocation matrix of UAVs $\boldsymbol{\rho}=\left\{\rho_i\right\}_{i\in\mathcal{I}}$. Denote the decision set as $\bm{Z}=\left\lbrace\bm{s},\bm{p},\boldsymbol{\beta},\boldsymbol{\rho}\right\rbrace$.
Then, the system \textcolor{blue}{CO} minimization problem is formulated as

\textbf{(P1):}
\begin{displaymath}
\min_{\bm{Z}}\mathcal{G}  \tag{20}
\end{displaymath}

\begin{subequations}\color{b}
\begin{align}
\mathrm{s.t.}\nonumber\\
&0 \leq a_i\leq a^{\max}, \forall i\in \mathcal{I}, \\
&s_i^m\in \left\lbrace0,1\right\rbrace, \forall i\in \mathcal{I}, \forall m\in \mathcal{M},\\
&\sum_{m=0}^Ms_i^m\leq 1, \forall i\in \mathcal{I},\\
&0\leq p_i\leq p_i^{\max}, \forall i\in \mathcal{I},\\
&0\leq \rho_i\leq \rho_i^{\max}, \forall i\in \mathcal{I},\\
&0 \leq \beta_i^m\leq 1,\forall i\in \mathcal{I}, \forall m\in \mathcal{M},\\
&\sum_{i\in \Lambda_m}\beta_i^m\leq 1, \forall m\in \mathcal{M},\\
&t_i^{loc} \leq T_i^{th}, \forall i\in \mathcal{I},\\
&t_i^{edge} \leq T_i^{th}, \forall i\in \mathcal{I}.
\end{align} 
\end{subequations}
In \textbf{(P1)}, constraint (20a) indicates the task arrivals is bounded by $a^{\max}$, constraints (20b) and (20c) denote the server selection decision constraints, respectively, constraint (20d) guarantees that the transmission power of UAV $i$ cannot exceed the maximum $p_i^{\max}$, constraint (20e) indicates the local CPU-cycle frequency of UAV $i$ is bounded by $\rho_i^{\max}$, constraint (20f) denotes that the computing resource allocation ratio is between 0 and 1, constraint (20g) ensures that the total computation resource allocated to UAVs at edge server $m$ is bounded by $F_m$, constraint (20h) and (20i) provide the maximum latency guarantee for local computing and edge computing, respectively.

\section{Problem Transformation and Solution Design}
\textcolor{b}{In \textbf{(P1)}, the variable $s_i^m$ is a binary integer variable that governs server selection decisions. Notably, the objective function incorporates terms involving the product of this binary variable and continuous variables, a formulation that introduces nonlinearity. Given the presence of both integer and continuous variables and nonlinear components, \textbf{(P1)} is identified as an MINLP. It is well-established that finding the optimal solution to an MINLP typically requires exponential time complexity, which reflects the inherent computational challenge of this type of optimization problems.}
Considering the large number of decision variables that scale exponentially with the number of UAVs and edge servers, we aim to devise a sub-optimal solution that boasts low complexity. 
We transform the original problem into four subproblems and employ a two-stage optimization that separately handles integer and continuous variables. In Stage I, we determine UAV-server selection decisions $\bm{s}$ through a matching game-based mechanism, detailed in Section IV-A. Subsequently, in Stage II, we derive the resource allocation parameters  $\bm{p}$, $\boldsymbol{\beta}$, and $\boldsymbol{\rho}$, given the Stage I solution.
Specifically, after each UAV determines which edge server to offload the tasks to, we introduce a transmission power allocation algorithm in Section IV-B. Regarding the computing resource allocation of the edge servers, a projected gradient descent based algorithm is devised in Section IV-C. Furthermore, the local computing resource allocation is optimized in Section IV-D. Finally, we propose a JCORA algorithm to optimize the total \textcolor{blue}{CO} of all the UAVs.
The entire solution framework for solving the formulated problems is illustrated in Fig. 2. 

\begin{figure}[htbp]
	\centerline{\includegraphics[width=0.35\textwidth]{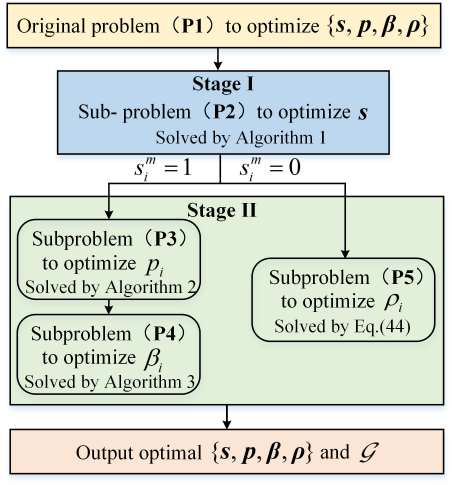}}
	\caption{The proposed solution framework for solving the formulated problems.}
	\label{fig}
\end{figure}

\subsection{ Server Selections of the UAVs: A Matching-Based Scheme}

We first assume that each UAV is assigned a default transmission power $p_i^{def}$. Then, the server selection problem is formulated as

\textbf{(P2):}
\begin{align}
	&\min_{\bm{s}}\mathcal{G}  \nonumber\\
	&s.t. \quad (20b), (20c).	
\end{align}
Since $s_i^m$ is a binary variable, it is difficult to solve the problem with traditional optimization methods. In the following, we opt to utilize the matching game method to tackle the problem.
We formulate \textbf{(P2)} as a two-sided matching problem, involving two distinct sets of agents: one set consisting of the UAVs, and the other set comprising the edge servers.
Since each server can serve multiple UAVs, yet one UAV can connect to only one edge server, we model \textbf{(P2)} as a many-to-one matching game.

\emph{a) Fundamentals:} We describe the matching as a triplet of $\left(\Theta,\Psi,\Phi \right)$:
\begin{itemize}
	\item[$\bullet$] 
	$\Theta=\left\lbrace \Theta_1, \Theta_2\right\rbrace $ denotes two disjoint sets of agents where $\Theta_1 \in \mathcal{I}$ is the set of UAVs that have not decided where to process tasks in time slot $t$, and $\Theta_2 \in \mathcal{M}$ is the set of the available edge servers.  
	\item[$\bullet$] 
	$\Psi=\left\lbrace \Psi_i^m, \Psi_m^i\right\rbrace$ denotes the preference lists of the UAVs and edge servers. Each $i\in \Theta_1$ has a descending ordered preferences on the servers, i.e., $\Psi_i^m=\left\lbrace m|m \in \Theta_2,m\succ_i m^{'} \right\rbrace $, where $\succ_i$ indicates the preference of UAV $i$ towards the servers. Similarly, each server has a descending ordered preference list over the UAVs, i.e., $\Psi_m^i=\left\lbrace i|i \in \Theta_1,i\succ_m i^{'}\right\rbrace $.
	\item[$\bullet$] 
	$\Phi \subseteq \left\lbrace i|i \in \mathcal{I}\right\rbrace \times \left\lbrace m|m \in \mathcal{M}\right\rbrace$ denotes the many-to-one matching between the UAVs and the edge servers. Specifically, each UAV is matched with at most one edge server, i.e., $\left| \Phi(i)\right|\leq 1, \forall i \in \mathcal{I}$, while each edge server is matched with multiple UAVs, i.e., $\Phi(m)\in \left\lbrace i|i \in \mathcal{I}\right\rbrace$.
\end{itemize}

\emph{b) Preference List:} The preference lists of each agent are constructed as follows. 

\begin{itemize}
	\item[$\bullet$] 
	For UAV $i$, it prefers edge servers with larger available channel capacity $R_i^m$ and larger residual computing capacity $F_m^{res}$. Therefore, the preference value for each UAV is expressed as
	\begin{equation}
		PL_i(m)=\xi_i^1R_i^m+\xi_i^2F_m^{res},
	\end{equation}
    where $\xi_i^1$	and $\xi_i^2$ indicate the weighted parameters. Then, we can construct the preference list for UAV $i$ by ranking the preference values as a descending order, yielding
   \begin{equation}
   	PL_i(m)>PL_i(m^{'})\Leftrightarrow m\succ_i m^{'},\Psi_i^m=\left\lbrace m, m^{'}\right\rbrace. 
   \end{equation}
    Furthermore, each UAV has a threshold of the preference value, i.e., $PL_i^{th}$, and UAV $i$ will process the tasks locally when  $PL_i(m)<PL_i^{th}, \forall m \in \mathcal{M}$.

	\item[$\bullet$] 
	For edge server $m$, it prefers UAVs with more tasks and larger tolerated delay\ccc{9714482}. Therefore, the preference list for each server is expressed as
	\begin{equation}
		PL_m(i)=\zeta_m^1a_i+\zeta_m^2T_i^{th},
	\end{equation}
	where $\zeta_m^1$ and $\zeta_m^2$ indicate the weighted parameters. Then, we can construct the preference list for edge server $m$  by ranking the preference values as a descending order, yielding
	\begin{equation}
		PL_m(i)>PL_m(i^{'})\Leftrightarrow i\succ_m i^{'},\Psi_m^i=\left\lbrace i, i^{'}\right\rbrace. 
	\end{equation}		
\end{itemize}

Based on the aforementioned analysis, we propose Algorithm 1 to illustrate the many-to-one matching between UAVs and edge servers.

\begin{algorithm}[htb] 
	\SetAlgoLined
	\caption{Many-to-One Matching Between UAVs and Edge Servers.}\label{algorithm}
	\KwIn{ The set of UAVs $\Theta_1\in\mathcal{I}$, the set of the available edge servers $\Theta_2\in\mathcal{M}$, the set of the default transmission power $\bm{p}^{def}$;}
	\KwOut{The optimal matching list $\Phi$, the optimal server selection strategy $\bm{s}$; }
	\textbf{Initialization:} $\Psi_i^m=\emptyset$,$\Psi_m^i=\emptyset$, $\Phi=\emptyset$, the rejected set $\Upsilon=\Theta_1$\;
	Calculate the preference value for UAVs and servers with Eq. (22) and Eq. (24), respectively\;
	Establish the preference list of UAVs and servers with Eq. (23) and Eq. (25), respectively\;
	\While{$\Upsilon\neq \emptyset$}{
	  \For{$i \in \Upsilon$}{
	    \eIf{$PL_i(m)<PL_i^{th}, \forall m\in M$}
	    {
	    	UAV $i$ processes the tasks locally\;
	    }
	    {UAV $i$ selects the most preferred server $m^{'}$ as its partner\;
	     Update the matching list of UAV $i$ as $\Phi(i)=\Phi(i)\cup m^{'}$ \;
	     Update the matching list of server $m^{'}$ as $\Phi(m^{'})=\Phi(m^{'})\cup i $ \;
	    }	 
	  }
	  \For{$m \in \Theta_2$}{
	  \eIf{$\sum_{i\in \mathcal{I}}s_i^m\leq \Lambda_m$ and server $m$ has enough computation resources}
	  {
	  	The server $m$ keeps all the matched UAVs and remove the matched UAVs from $\Upsilon$\;
	  }
	  {Update the matching list of server $m$ as $\Phi(m)=\Phi(m)\backslash \Upsilon^{rej}$, where $\Upsilon^{rej}$ is the set of less preferred UAVs \;
	   Update $\Upsilon$	as $\Upsilon=\Upsilon\cup \Upsilon^{rej}$ \;
	  	\For{$i\in \Upsilon^{rej}$}{
	  	  Update the preference list of UAV $i$ as $\Psi_i^m=\Psi_i^m \backslash m$ \;
	  	  Update the matching list  of UAV $i$ as $\Phi(i)=\Phi(i)\backslash m$ \;
	  	}	  		  	
	   }
	  }	
	}
		
\end{algorithm}

In particular, the matching should be stable, the blocking pair in many-to-one matching is defined as follows. 

\emph{Definition 1:} Blocking Pair in Many-to-One Matching: A pair $\left(i,m\right) $ is deemed as a blocking pair, if: 1) UAV $i$ is unserved or UAV $i$ prefers server $m$ to its current matching $\Phi(i)$ and 2) server $m$ is underutilized or prefers UAV $i$ to at least one UAV in existing matching $\Phi(m)$.

\emph{Definition 2:} A matching is stable if and only if there are no blocking pairs.

\emph{Theorem 1.} The optimal matching $\Phi$ in Algorithm 1 is stable for all $i \in \mathcal{I}$ and $m \in \mathcal{M}$.

\emph{Proof:}

We assume that matching $\Phi$ is unstable, i.e., there is one unmatched pair $i \in \mathcal{I}$ and $m \in \mathcal{M}$ that prefer each other than their current matching partners, i.e., $i\not\in \Phi(m)$ and $m\not\in \Phi(i)$. Then, the following conditions are satisfied:
\begin{subequations}
	\begin{align}
		&m\succ_i \Phi(i),  \\
		&i\succ_m i^{'}, i^{'}\in \Phi(m)\\
		&i\not\in \Phi(m),\\
		&m\not\in \Phi(i).
	\end{align} 
\end{subequations}
We prove the theorem by showing that the necessary conditions in (26) cannot be satisfied simultaneously\ccc{10024868}. If (26a) holds true, we deduce that UAV $i$ prefers server $m$ to its current matching $\Phi(i)$. However, condition (26c) indicates that UAV $i$ is less preferred by server $m$ than any $i^{'}\in \Phi(m)$, i.e., $i^{'}\succ_m i, i^{'}\in \Phi(m)$, which contradicts the condition in (26b). In this case, Although UAV $i$ prefers server $m$, server $m$ does not prefer UAV $i$. Therefore, the blocking pair cannot be formed between $i \in \mathcal{I}$ and $m \in \mathcal{M}$. Then, the matching obtained from Algorithm 1 is stable.

\subsection{Optimization of Transmission Power Allocation}
Once UAV $i$ identifies the appropriate MASS or HAP for task offloading, we can proceed to optimize the transmission power in accordance with their specific requirements, thereby minimizing the \textcolor{blue}{CO}.
The transmission power allocation problem is modeled as

\textbf{(P3):}
\begin{align}
	\min_{\bm{p}}&\sum_{i \in \mathcal{I}}\left(\mu_i^t\cdot\frac{a_iD}{R_i^m}+ \mu_i^e\cdot\frac{p_ia_iD}{R_i^m}\right)  \nonumber\\
	&s.t. \quad (20d).	
\end{align}
Substituting (5) into (27), \textbf{(P3)} is reformulated as

\textbf{(P3.1):}

\begin{align}
	\min_{\bm{p}}&\sum_{i \in \mathcal{I}}\frac{a_iD}{W_i^m}\cdot\frac{\mu_i^t+\mu_i^ep_i}{\log_2\left(1+\frac{p_iH_i^m}{\sigma^2} \right) }  \nonumber\\
	&s.t. \quad (20d).	
\end{align}
For convenience, let $B_i=\frac{a_iD}{W_i^m}$, $C_i=H_i^m/\sigma^2$.
We introduce $\Xi(p_i)$ to solve \textbf{(P3.1)} and define it as
\begin{equation}
  \Xi(p_i)=B_i\cdot\frac{\mu_i^t+\mu_i^ep_i}{\log_2\left(1+C_ip_i \right) }. 	
\end{equation}
\textbf{(P3.1)} is non-convex as the second-order derivative of $\Xi(p_i)$ is not always positive. In view of this, we employ the quasi-convex optimization technique to address problem (28) based on the following lemma.

\emph{Lemma 1:} $\Xi(p_i)$ is quasi-convex in the domain, and it satisfies the quasi-convex second-order condition and there is a point $p_i^0$ satisfying $\Xi^{'}(p_i)=0$ and $\Xi^{''}(p_i)>0$.

\emph{Proof:}

Obviously, $\Xi(p_i)$ is twice differentiable on $\mathbb{R}$, The first-order derivative of $\Xi(p_i)$ is given by
\begin{equation}
\frac{\partial\Xi(p_i)}{\partial p_i}=B_i\frac{\mu_i^e\log_2\left(1+C_ip_i \right)-\frac{C_i\left( \mu_i^t+\mu_i^ep_i\right) }{\ln2\left(1+C_ip_i \right)}} {\left[\log_2\left(1+C_ip_i \right)\right]^2 }.
\end{equation}

When $\Xi^{'}(p_i^0)=0$, we obtain
\begin{equation}
	\Omega(p_i^0)=\mu_i^e\log_2\left(1+C_ip_i^0 \right)-\frac{C_i\left( \mu_i^t+\mu_i^ep_i^0\right) }{\ln2\left(1+C_ip_i^0 \right)}=0.
\end{equation}
The second-order derivative of $\Xi(p_i)$ is given at the top of this page.

\begin{figure*}[ht] %hb代表放在文章底部，%ht为放在文章顶部 
	\centering
	\begin{equation}
		\frac{\partial^2\Xi(p_i)}{\partial^2 p_i}= B_i\frac{\ln2C_i\log_2\left(1+C_ip_i \right)\left[C_i\left( \mu_i^t+\mu_i^ep_i\right)-2\mu_i^e\left(1+C_ip_i \right) \right]+2C_i^2\left( \mu_i^t+\mu_i^ep_i\right)}{\left[\ln2\left(1+C_ip_i \right) \right]^2\log_2^3\left(1+C_ip_i \right) }.  	
	\end{equation}
		
\end{figure*}

Substituting $p_i^0$ into the second-order derivative of $\Xi(p_i)$, we obtain 
\begin{equation}
	\frac{\partial^2\Xi(p_i^0)}{\partial^2 p_i^0}=B_i\frac{C_i^3\left( \mu_i^t+\mu_i^ep_i^0\right)^2}{(\ln2)^2\mu_i^e\left(1+C_ip_i^0 \right)^3\log_2^3\left(1+C_ip_i^0 \right)}\geq0.
\end{equation}
Thus, $\Xi(p_i)$ is quasi-convex in the domain $\left[0,p_i^{\max}\right] $.

Note that a quasi-convex function reaches a local optimum at the point of decreasing first-order derivatives to 0. Any local optimum of a quasi-convex function is a global optimum\ccc{10376220}. Since $\Omega(0)=-\frac{C_i\mu_i^t}{\ln2}<0$, we obtain $\Xi^{'}(0)<0$ and $\Xi^{'}(p_i)$ is a monotonically increasing function and is negative at the starting point $p_i=0$. Therefore, based on Lemma 1, we confirm that the optimal solution $p_i^*$ either lies at the constraint border, i.e., 
$p_i^*=p_i^{\max}$, or satisfies $p_i^*=p_i^0$, where $\Xi^{'}(p_i^0)=0$. Then, the optimal $p_i^*$ is expressed as
\begin{equation}
	p_i^*=\min\left\lbrace p_i^0,p_i^{\max}\right\rbrace.
\end{equation}

Based on the above analysis, we design a transmission power allocation (TPA) algorithm based on the low-complexity bisection method to evaluate $\Xi^{'}(p_i)$ in each iteration, so as to obtain the optimal $p_i^*$, as shown in Algorithm 2.

\begin{algorithm}
	\SetAlgoLined
	\caption{TPA Algorithm.}\label{algorithm}
	\KwIn{ The set of UAVs $\mathcal{I}$, the set of edge servers $\mathcal{M}$;}
	\KwOut{The optimal transmission power allocation strategy $\bm{p}$; }
	
	\textbf{Initialization:} The optimality tolerance parameter $\delta$\;
	Calculate $\Omega(p_i^{\max})$ with Eq. (31)\;
	\eIf{$\Omega(p_i^{\max})\leq 0$}{
	   Set $p_i^*=p_i^{\max}$ \;
	}{
	  Set the lower bound as $p_i^L=0$ and the upper bound as $p_i^U=p_i^{\max}$ \;
	  \While{$p_i^U-p_i^L>\delta$}{
	    Calculate the current value of $p_i^{cur}=\left(p_i^U+p_i^L\right) /2$ \;
	    \eIf{$\Omega(p_i^{cur})\leq 0$}{
	       Set $p_i^L=p_i^{cur}$ \;
	    }{
	       Set $p_i^U=p_i^{cur}$ \;
	    }
	  
	  }
	  Set $p_i^*=\left(p_i^U+p_i^L\right) /2$ \;
	}
	
\end{algorithm}

\subsection{Optimization of Edge Computing Resource Allocation }
After UAV $i$ successfully offloads tasks to the designated edge server, we will optimize the allocation of edge computing resources to minimize the \textcolor{blue}{CO}, formulated as
 
\textbf{(P4):}

\begin{align}
	\min_{\boldsymbol{\beta}}&\sum_{i\in \mathcal{I}} \left\lbrace 
	\mu_i^t\cdot\frac{a_iD\alpha}{\beta_i^mF_m}+\mu_i^e\cdot\varepsilon_m\left(\beta_i^mF_m\right) ^2a_iD\alpha\right\rbrace  \nonumber\\
	&s.t. \quad (20f), (20g).	
\end{align}

Notice that the constraints in (20f) and (20g) are convex and the second-order derivative of the objective function is greater than zero, \textbf{(P4)} is a convex problem. We employ the projected Gradient descent method to find the optimal $\beta_i^{m*}$. For convenience, let
\begin{equation}
  \mathcal{L}(\boldsymbol{\beta})=\sum_{i\in \mathcal{I}} \left\lbrace 
  \mu_i^t\cdot\frac{a_iD\alpha}{\beta_i^mF_m}+\mu_i^e\cdot\varepsilon_m\left(\beta_i^mF_m\right) ^2a_iD\alpha\right\rbrace.
\end{equation}

The gradient of $\mathcal{L}(\boldsymbol{\beta})$ is denoted as $\triangledown\mathcal{L}=\left\lbrace \frac{\partial\mathcal{L}}{\partial\beta_i^m}\right\rbrace $, where
\begin{equation}
	\frac{\partial\mathcal{L}}{\partial\beta_i^m}=2\mu_i^e\varepsilon_mF_m^2a_iD\alpha\beta_i^m-\frac{\mu_i^ta_iD\alpha}{F_m\left(\beta_i^m\right)^2}.
\end{equation}
The update rule of gradient descent method is denoted as
\begin{equation}
	\left(\beta_i^m\right)^{k+1}=\left(\beta_i^m\right)^{k}-\eta\frac{\partial\mathcal{L}}{\partial\beta_i^m}\mid_{\beta_i^m=\left(\beta_i^m\right)^{k}},
\end{equation}
where $\left(\beta_i^m\right)^{k}$ is the value of $\beta_i^m$ at the $k$-th iteration, $\eta$ is the learning rate (step size), which needs to be manually set.

Subsequently, we will employ a projection operation to ensure that the value of $\beta_i^m$ remains within the bounds defined by the constraint conditions.

\begin{itemize}
\item[$\bullet$] 
 Projection to the $[0,1]$ interval: We first need to perform constraint projection on each $\beta_i^m$ to ensure that $0\leq\beta_i^m\leq1$, yielding:
 \begin{equation}
 	\left(\beta_i^m\right)^{k+1}=\max\left\lbrace0,\min\left[1,\left(\beta_i^m\right)^{k+1}\right]  \right\rbrace .
 \end{equation}
   
\item[$\bullet$] 
Projection to $\sum_{i\in \Lambda_m}\beta_i^m\leq 1$: After each update, we check the value of $\sum_{i\in \Lambda_m}\beta_i^m$, if $\sum_{i\in \Lambda_m}\beta_i^m\geq 1$, we need to scale each $\beta_i^m$ so that the sum	$\sum_{i\in \Lambda_m}\beta_i^m$ does not exceed $1$. Let $scale=\frac{1}{\sum_{i\in \Lambda_m}\left(\beta_i^m\right)^{k+1}}$, then, each $\left(\beta_i^m\right)^{k+1}$ is scaled by
 \begin{equation}
	\left(\beta_i^m\right)^{k+1}=\left(\beta_i^m\right)^{k+1}\times scale.
\end{equation}

\end{itemize}

Driving from the above analysis, we propose a computation resource allocation (CRA) algorithm base on the projected Gradient descent method to obtain the optimal $\beta_i^{m*}$ for each UAV, as shown in Algorithm 3, \textcolor{b}{where clip$(x, x_{\min}, x_{\max}) = \max\left[x_{\min}, \min(x, x_{\max})\right]$ ensures the solution remains within the feasible domain by constraining values between a lower bound and an upper bound}. 

\begin{algorithm}
	\SetAlgoLined
	\caption{CRA Algorithm.}\label{algorithm}
	\KwIn{ Learning rate $\eta$, convergence tolerance $\epsilon$, maximum iterations $K$;}
	\KwOut{The optimal solution $\boldsymbol{\beta}^*$; }
	
	\textbf{Initialization:} $0\gets k$, set each $\beta_i^m$ with an initial value $\beta_i^{m0}$\;
	\While{$k \leq K$}{
	   Calculate gradient $\triangledown\mathcal{L}$ with Eq. (37)\;
	   \For{$i\in \mathcal{I}$}{
	      Update $\left(\beta_i^m\right)^{k}$ with Eq. (38)\;
	   }
	   Project onto feasible region: $\left(\beta_i^m\right)^{k} \gets \text{clip}\left[\left(\beta_i^m\right)^{k},0,1  \right] $\;
	   \If{$\sum_{i\in \Lambda_m}\beta_i^m\geq 1$}{
	      Update $\left(\beta_i^m\right)^{k}$ with Eq. (40)\;
	   }
	   \If{$\Vert\left(\beta_i^m\right)^{k+1}-	\left(\beta_i^m\right)^{k}\Vert<\epsilon$}{
	       break  \;
	   }
	   Update $\beta_i^m=\left(\beta_i^m\right)^{k+1}$\;
	}
	
\end{algorithm}

\subsection{Optimization of Local Computation Resource Allocation}
When $s_i^m=0$, the tasks of UAV $i$ will be executed locally. By selecting optimal processing parameters, the \textcolor{blue}{CO} can be effectively minimized. Following this, we will address the optimization of local computing resource allocation, formulated as:

\textbf{(P5):}
\begin{align}
	\min_{\boldsymbol{\rho}}&\sum_{i\in \mathcal{I}}\left(\mu_i^t\cdot\frac{a_iD\alpha}{\rho_i}+\mu_i^e\cdot\varepsilon_i\rho_i^2a_iD\alpha\right)   \nonumber\\
	&s.t. \quad (20e), (20h).	
\end{align}

\textbf{(P5)} is a convex optimization problem regarding $\boldsymbol{\rho}$ and decoupled with respect to each UAV. Its solution is obtained at the poles or at the boundary. We first obtain its pole value $\rho_i^0$ by assigning its first-order derivative function to zero. Then, we get  
\begin{equation}
	\rho_i^0=\sqrt[3]{\frac{\mu_i^t}{2\mu_i^e\varepsilon_i}}.
\end{equation}
According to constraint (20h), we have
\begin{equation}
	\rho_i\geq \frac{a_iD\alpha}{T_i^{th}}=\rho_i^{\min}.
\end{equation}
With Eq. (42) and  Eq. (43), we obtain the optimal $\rho_i^*$ as
\begin{equation}
	\rho^*=
	\begin{cases}
		\rho_i^{\min}, \rho_i^0<\rho_i^{\min}\\
		\rho_i^0, \rho_i^{\min}\leq\rho_i^0\leq\rho_i^{\max}\\
		\rho_i^{\max},\rho_i^0>\rho_i^{\max}.
	\end{cases}
\end{equation}

Based on the aforementioned analysis, we propose the JCORA algorithm to jointly optimize server selection, transmission power allocation, edge computing resource allocation, and local computing resource allocation, as outlined in Algorithm 4.

\begin{algorithm}
	\SetAlgoLined
	\caption{JCORA Algorithm.}\label{algorithm}
	\KwIn{ The set of UAVs $\mathcal{I}$, the set of edge servers $\mathcal{M}$, the default transmission power $\bm{p}^{def}$, the optimality tolerance $\delta$, learning rate $\eta$, convergence tolerance $\epsilon$, maximum iterations $K$;}
	\KwOut{the optimal $\bm{s}^*$, $\bm{p}^*$, $\boldsymbol{\beta}^*$, $\boldsymbol{\rho}^*$ and the optimal CO $\mathcal{G}^*$ ; }
	 Obtain $\bm{s}^*$ with Algorithm 1\;
	\For{$i\in \mathcal{I}$}{	  
	   \eIf{$s_i^m=0$}{
	   	Calculate $\rho_i^*$ with Eq. (44) \;
	   	}{
	   	Obtain $p_i^*$ with Algorithm 2 \;
	   	Obtain $\beta_i^{m*}$ with Algorithm 3 \;
	   	}
	   Calculate $\mathcal{G}_i^*$ for each UAV \;
	}
	
\end{algorithm}

Finally, we analyze the computational complexity of the proposed
algorithms. For Algorithm 1, in the worst
case, any UAV could be rejected at most $\mathcal{M}$ times, i.e., the UAV is rejected by all edge servers. Whenever a UAV is rejected by its preferred server, in the subsequent iteration, there are at most $\min\left\lbrace \mathcal{I},\mathcal{M} \right\rbrace$ servers that need to update their preference lists.
Therefore, the
worst-case complexity of Algorithm 1 is $\mathcal{O}(\mathcal{M}\cdot(2\mathcal{I}+\min\left\lbrace \mathcal{I},\mathcal{M} \right\rbrace ))$, where $\mathcal{I}$ and $\mathcal{M}$ are the number of UAVs and edge servers, respectively.
We denote the number of iterations of Algorithm 2 and 3 as $Q$ and $K$, respectively. Then, we obtain the computation complexity of Algorithm 2 and 3 as $\mathcal{O}(\log_2Q)$  and $\mathcal{O}(K)$ for each UAV, respectively. Therefore, the total computational complexity of the proposed algorithms is $\mathcal{O}(\mathcal{M}\cdot(2\mathcal{I}+\min\left\lbrace \mathcal{I},\mathcal{M} \right\rbrace )+\mathcal{I}(\log_2Q+K))$.
We see that the proposed algorithms have a low
complexity compared to conventional solutions, e.g., the exhaustive search  policy with a complexity of $\mathcal{O}(\mathcal{M}^\mathcal{I})$.

\section{Performance Evaluation}
In this section, we conduct simulations to verify the effectiveness of the proposed algorithms. More specifically, we assess the impact of key parameters and compare our proposed scheme with the following benchmark schemes.

\begin{itemize}
%\item[$(a)$] 
%Pure local computing (PLC): In this scheme, All the MASSs process their tasks locally without task offloading, i.e., $s_{mn}=0$, $\forall m\in \mathcal{M},\forall n\in \mathcal{N}$. 
\item[$(a)$] 
Local Computing (LC): All UAVs choose to perform the tasks locally, i.e., $s_i^m = 0, \forall i \in \mathcal{I}$. 
\item[$(b)$] 
Only Match Game (OMG)\ccc{9696188}: This scheme determines task offloading through a match game without incorporating joint resource allocation. Each UAV utilizes a constant transmission power, whereas the edge servers adopt an equitable sharing strategy for computing resource allocation.
\item[$(c)$] 
Distributed edge user allocation (DEUA) scheme: This scheme is extended from\ccc{8823046} to be applied to our model. When the UAVs compete for the matched edge servers, the winner is selected randomly. 
\end{itemize}

 \subsection{System setup}

We conduct all the simulations using MATLAB on a PC configured with a Core i7-10510U 1.80 GHz CPU and 8 GB of RAM. 
We consider an aerial-marine integrated network comprised of one HAPS and four MASSs which navigate autonomously on the ocean. We assume that all UAVs are randomly distributed within the service
range of the HAPS. 
The maximum number of task arrivals of each UAV is $10$, and the size of each task is $0.5$ Mbits.
Each UAV communicates with the corresponding MASS  via OFDMA, utilizing a channel bandwidth of $10$ MHz, and each UAV communicates with the HAPS through OFDMA, employing a channel bandwidth of $10$ MHz. The main parameters are shown in Table II.

\begin{table}[htbp]
\centering
\caption{Simulation Parameter Settings}
\begin{tabular}{l|l}%l=left, r=right,c=center分别代表左对齐，右对齐和居中，字母的个数代表列数
\hline
\hline
Parameters &Values  \\ 
\hline
Carrier frequency ($f_i$) &$0.2$ GHz  \\
\hline
Speed of light ($c$) &$3 \times 10^{8}$ m/s  \\
\hline
Transmission bandwidth of MASS  &$10$ MHz  \\ 
\hline
Transmission bandwidth of HAPS  &$10$ MHz  \\
 \hline
Background noise power ($\sigma^2$)  &$1\times 10^{-13}$ W  \\
\hline
Path-loss exponent ($\lambda$)  &$1.5$   \\
\hline
Height of HAPS ($Z_H$)  &$20$km  \\
\hline
Height of UAV $i$ ($Z_i$)  &$200$m  \\
\hline
Link loss ($\delta_i^{LOS}$ and $\delta_i^{NLOS}$ )\ccc{10373079}  &$0.1, 21$   \\
\hline
Environmental parameters ( $\chi_1$ and $\chi_2$)\ccc{10373079}  &$4.88,0.43$s  \\
\hline
Learning rate ($\eta$)  &$0.01$  \\
\hline
Number of CPU cycles for processing one bit of \\
data ($\alpha$) &\textcolor{b}{$1 \times10^3$}\\ 
\hline
Maximum computing capability of UAVs ($\rho_i^{\max}$)  &\textcolor{b}{$5 \times 10^{9}$} cycles/s \\
\hline
\hline
\end{tabular}
\end{table}

 \subsection{Performance evaluation}
Fig. 3 depicts the total energy consumption and the overall latency for completing all UAVs' tasks associated with different task sizes, respectively, as we adjust the UAVs’ preference for time, $\mu_i^t$, from $0.1$ to $0.9$ while changing the UAVs’ preference to energy accordingly as $\mu_i^e=1-\mu_i^t$, $\forall i\in \mathcal{I}$. 
We observe that an increase in $\mu_i^t$ leads to a reduction in the overall latency, albeit at the expense of elevated energy expenditure.
Moreover, with the expansion of task size, a corresponding increase was observed in both energy consumption and the overall latency.
The observed trend stems from inherent system limitations, as the increased computational demands from the escalating task size of UAVs necessitates greater temporal and energy resources for task execution.

\begin{figure}[htbp]
	\begin{minipage}{0.49\linewidth}
		\vspace{3pt}
		\centerline{\includegraphics[width=\textwidth]{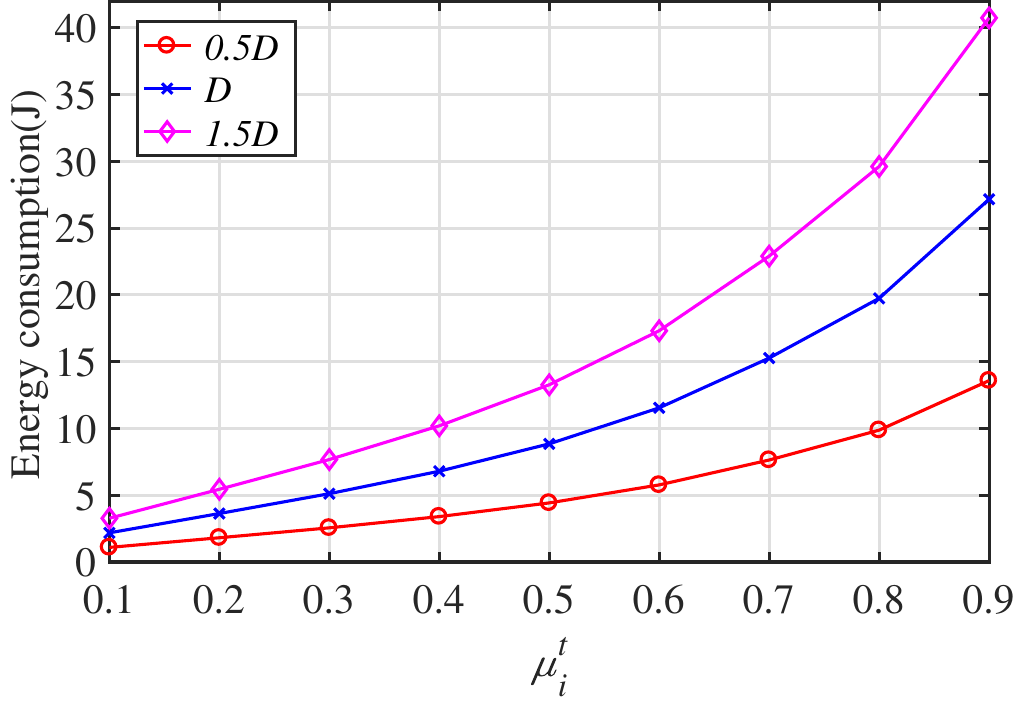}}
	 
		\centerline{(a) Energy consumption vs. $\mu_i^t$}
	\end{minipage}
	\begin{minipage}{0.49\linewidth}
		\vspace{3pt}
		\centerline{\includegraphics[width=\textwidth]{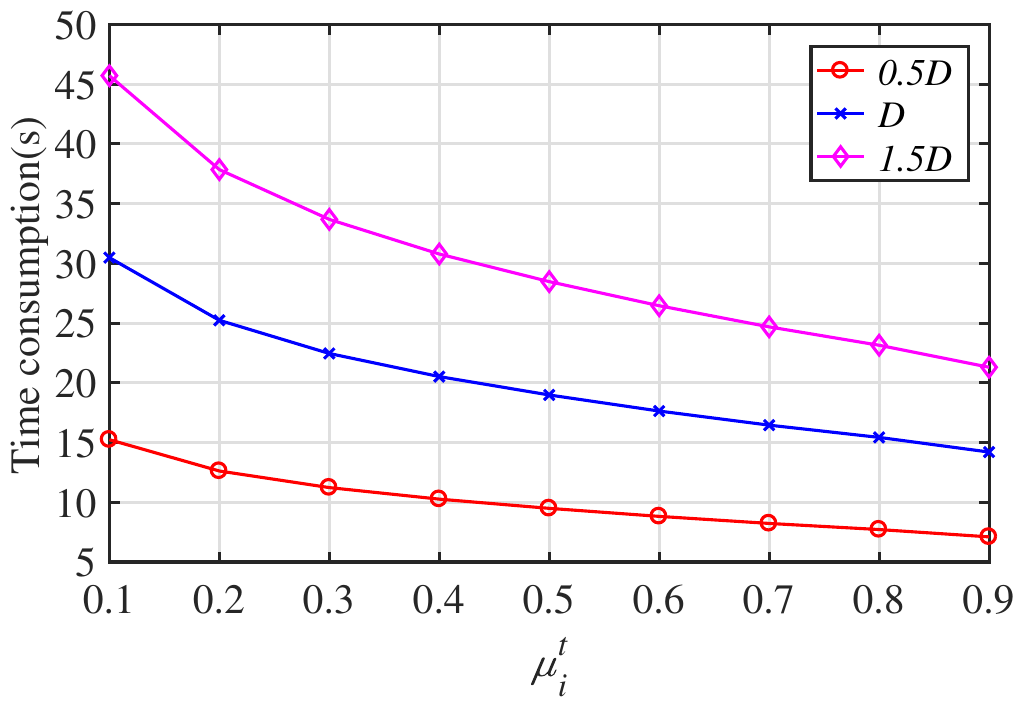}}
	 
		\centerline{(b) Time consumption vs. $\mu_i^t$}
	\end{minipage}
 
	\caption{The total energy consumption and overall latency associated with different task sizes under different UAV preferences for time.}
	\label{fig}
\end{figure}

Fig. 4 illustrates the total energy consumption and the overall latency for completing all UAVs' tasks associated with different number of UAVs, respectively, when we adjust the UAVs’ preference for time, $\mu_i^t$, from $0.1$ to $0.9$ while changing the UAVs’ preference to energy accordingly as $\mu_i^e=1-\mu_i^t$, $\forall i\in \mathcal{I}$. 
Consistent with the trends observed in Fig. 3, we observe that increasing $\mu_i^t$ effectively reduces system latency, though this performance enhancement comes at the cost of increased energy consumption.
Furthermore, an increase in the number of UAVs results in a corresponding surge in both energy consumption and overall latency. This is attributable to the heightened competition among UAVs for limited resources, wherein the increasing number of UAVs reduces the offloading efficiency and creates computational bottlenecks.

 \begin{figure}[htbp]
 	\begin{minipage}{0.49\linewidth}
 		\vspace{3pt}
 		\centerline{\includegraphics[width=\textwidth]{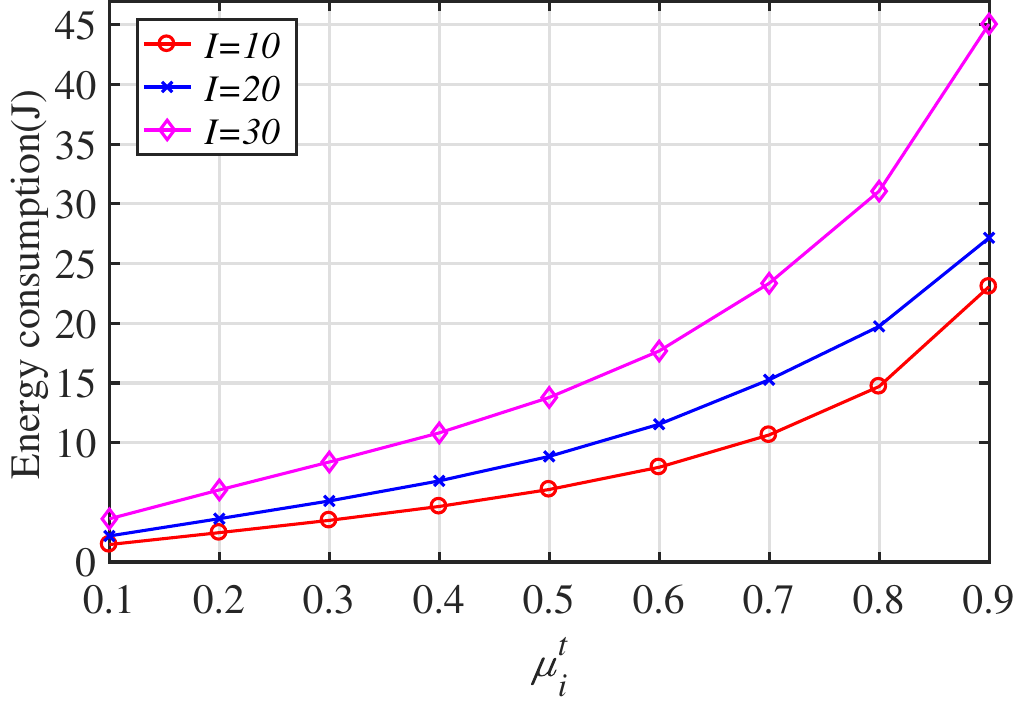}}
 		
 		\centerline{(a) Energy consumption vs. $\mu_i^t$}
 	\end{minipage}
 	\begin{minipage}{0.49\linewidth}
 		\vspace{3pt}
 		\centerline{\includegraphics[width=\textwidth]{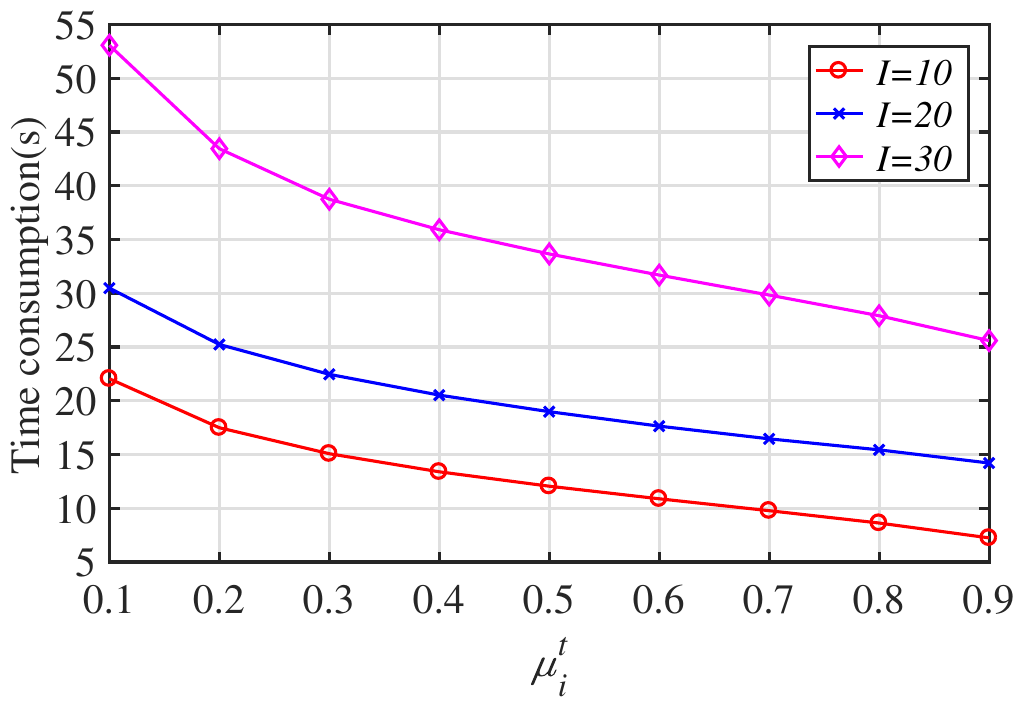}}
 		
 		\centerline{(b) Time consumption vs. $\mu_i^t$}
 	\end{minipage}
 	
 	\caption{The total energy consumption and overall latency associated with different number of UAVs under different UAV preferences for time.}
 	\label{fig}
 \end{figure}

Fig. 5 illustrates the impact of task quantity $a_i$ and the number of UAVs on the \textcolor{blue}{CO}, i.e., the overall latency and the energy consumption, respectively. 
Specifically, Fig. 5(a) illustrates that system performance degrades with increasing values of $a_i$, manifesting in elevated latency and amplified energy consumption. This phenomenon originates from the cumulative resource requirements of the system, where the growing workload necessitates additional processing capabilities from MASSs, UAVs, and HAPS, thereby intensifying both temporal and energy expenditures.
Fig. 5(b) indicates that both the latency and energy consumption exhibit progressive deterioration with increasing UAV numbers, due to the inherent computational demands, where expanding UAV numbers generate greater processing workloads that demand additional time and energy resources.

\begin{figure}[htbp]
	\begin{minipage}{0.49\linewidth}
		\vspace{3pt}
		\centerline{\includegraphics[width=\textwidth]{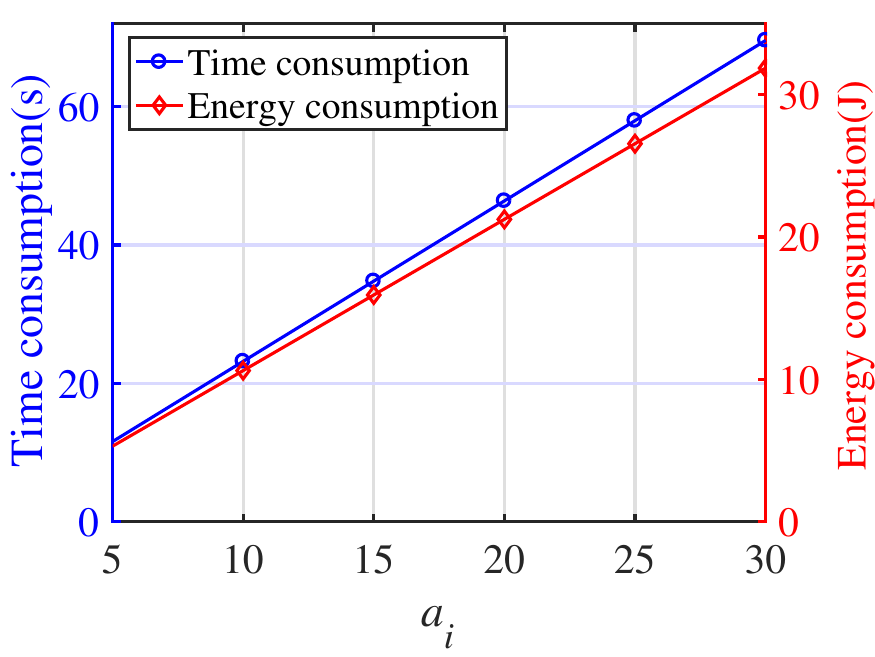}}
		
		\centerline{(a) CO vs. $a_i$}
	\end{minipage}
	\begin{minipage}{0.49\linewidth}
		\vspace{3pt}
		\centerline{\includegraphics[width=\textwidth]{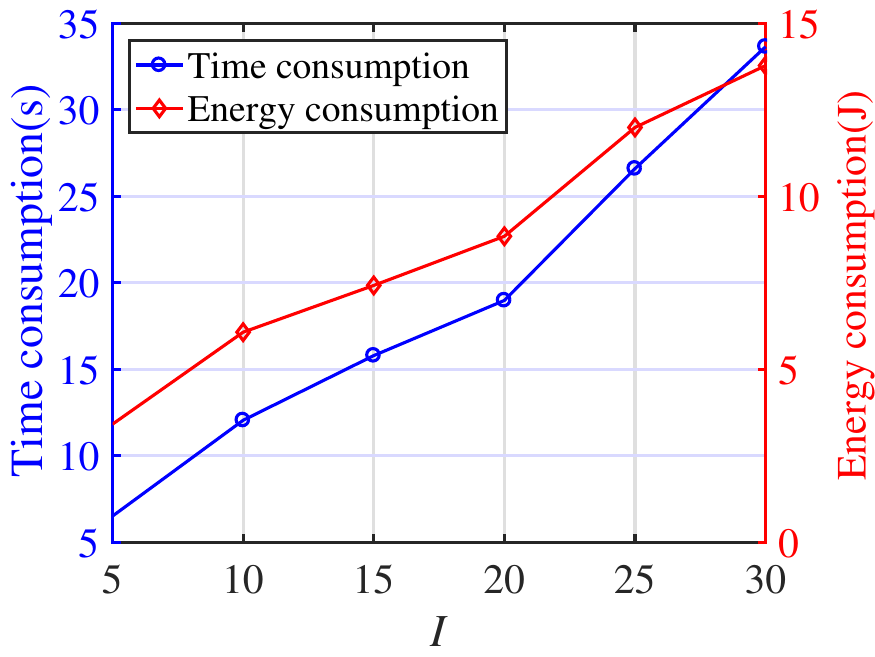}}
		
		\centerline{(b) CO vs. $I$}
	\end{minipage}
	
	\caption{The \textcolor{blue}{CO} associated with different number of tasks and number of UAVs, respectively.}
	\label{fig}
\end{figure}

Fig. 6 illustrates the performance comparison of the proposed scheme with other three benchmarks under different parameters (i.e., task quantity, the number of UAVs, the UAV preferences for time and the computational capability of MASSs). Specifically, in Fig. 6(d), $\omega$ indicates the multiple of MASSs' computing resources $F_m$, namely 0.5$F_m$, 1.0$F_m$, 1.5$F_m$, 2.0$F_m$ and 2.5$F_m$.
Figs. 6(a), 6(b), 6(c) and 6(d) clearly demonstrate the superiority of our proposed scheme, which effectively diminishes the overall \textcolor{blue}{CO}. This is attributed to our method of determining optimal task offloading decisions and resource allocation strategies by minimizing the total \textcolor{blue}{CO}.

\begin{figure}[htbp]	
	\begin{minipage}{0.5\linewidth}%可修改0.32为其他比例，调整大小
			\vspace{1pt}
			%这个图片路径替换成你的图片路径即可使用
			\centerline{\includegraphics[width=\textwidth]{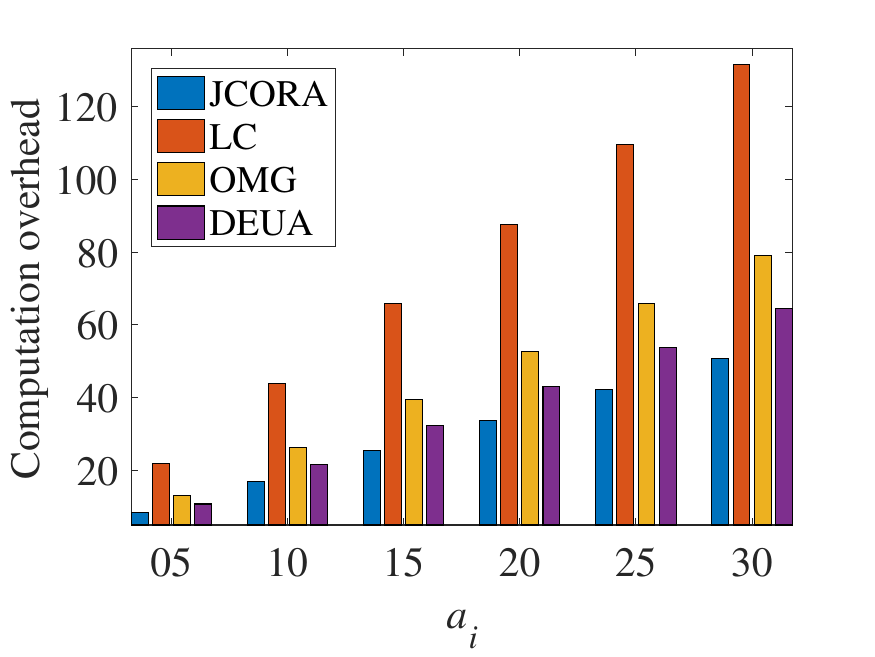}}
			% 加入对这列的图片说明
			\centerline{(a) CO vs. $a_i$ }
		\end{minipage}
	\begin{minipage}{0.5\linewidth}
			\vspace{1pt}
			\centerline{\includegraphics[width=\textwidth]{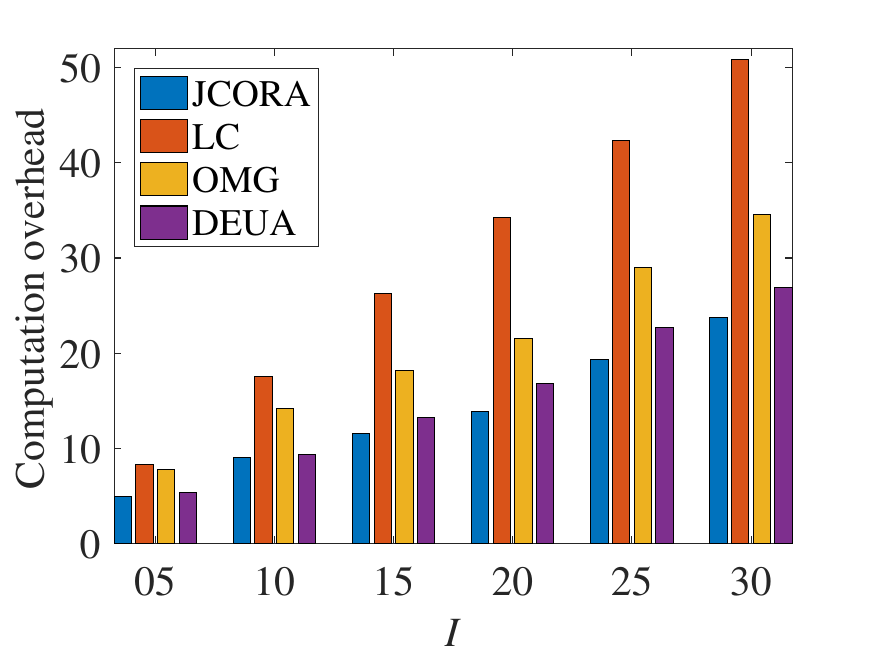}}
			
			\centerline{(b) CO vs. $I$}
		\end{minipage}
	\begin{minipage}{0.5\linewidth}
			\vspace{1pt}
			\centerline{\includegraphics[width=\textwidth]{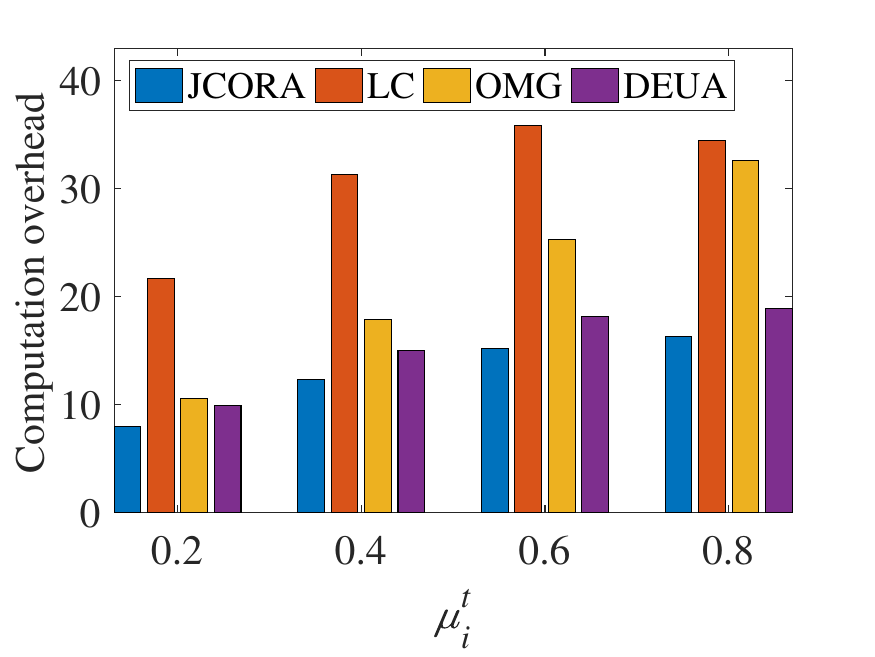}}
			
			\centerline{(c) CO vs. $\mu_i^t$}
		\end{minipage}
	\begin{minipage}{0.5\linewidth}
			\vspace{1pt}
			\centerline{\includegraphics[width=\textwidth]{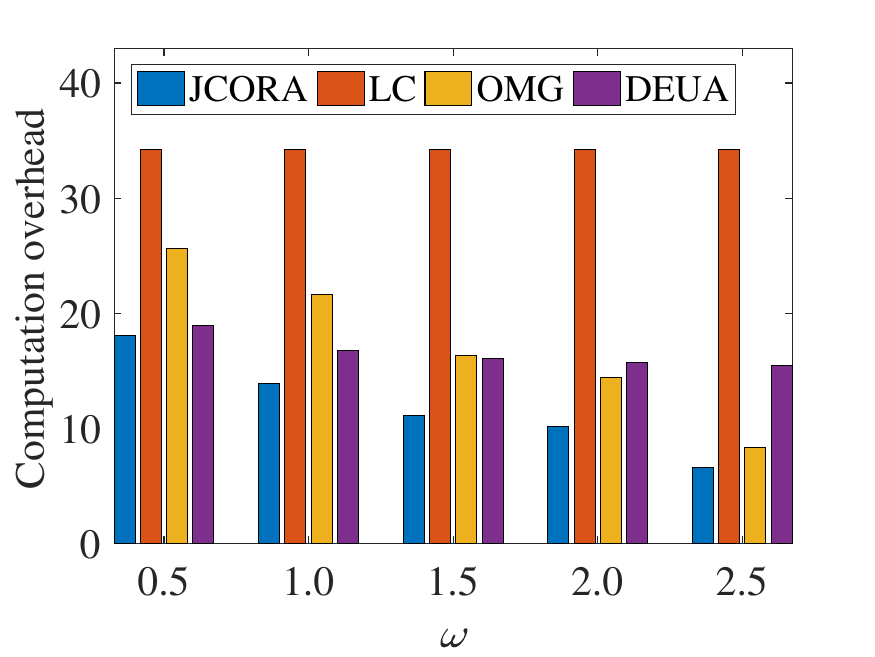}}
			
			\centerline{(d) CO vs. $\omega$}
		\end{minipage}

	\caption{Performance comparison of our proposed scheme with other benchmarks under different parameters (i.e., $a_i$, $I$, $\mu_i^t$ and $\omega$).}
\label{fig}
\end{figure}

\section{Conclusion and Future Work}

In this paper, we have considered an MSAR scenario and proposed a joint task offloading and resource allocation scheme for surveillance UAVs, by optimizing the server selection, transmission power allocation, edge computing resource allocation, and local computing resource allocation. The objective is to minimize the system total \textcolor{blue}{CO}, quantified by a weighted-sum of task completion time and energy consumption of each UAV.
We define a novel scenario where both MASSs and the HAPS are equipped with edge-servers to provide computing services for surveillance UAVs and the computation workloads of surveillance UAVs can be offloaded to MASSs and the HAPS via a multi-access approach. 
To address the stochastic UAV task arrivals and dynamic marine channel conditions, we decompose the original problem into four subproblems via a two-stage optimization: (1) edge server selection through many-to-one matching, (2) UAV transmission power allocation via quasi-convex optimization, (3) edge computing resource distribution using projected gradient descent, and (4) local computation resource assignment through convex optimization. Finally, we propose a JCORA algorithm aimed at minimizing the overall \textcolor{blue}{CO} of the system.
Simulation results demonstrate that our scheme outperforms benchmark schemes, highlighting its potential to significantly improve the efficiency of MSAR operations.
For future work, we will consider the integration of advanced artificial intelligence (AI)-driven techniques to further enhance the decision-making process in task offloading and resource allocation for MSAR.

% use section* for acknowledgment
\section*{Acknowledgment}
This work was supported by the National Natural Science Foundation of China (No. 62371085 and No.51939001) and the Scientific research fund of Liaoning Provincial Department of Education
(LJ212513631002).

\ifCLASSOPTIONcaptionsoff
  \newpage
\fi

\bibliographystyle{ieeetr}
\bibliography{Reference}

% that's all folks
\end{document}